\documentclass[a4paper]{jpconf}

\usepackage{./aas_macros}
\usepackage{graphicx}
\usepackage{amssymb}

\begin{document}
%opening
\title{Magneto-elastic torsional oscillations of magnetars}
\author{Michael Gabler$^{1,2,3}$
, Pablo Cerd\'a-Dur\'an$^1$, Jos\'e A.~Font$^2$ , Ewald M\"uller$^1$ and Nikolaos Stergioulas$^{3}$}
\address{$^1$ Max-Planck-Institut f\"ur Astrophysik,
  Karl-Schwarzschild-Str.~1, 85741 Garching, Germany}
\address{$^{2}$ Departamento de Astronom\'{\i}a y Astrof\'{\i}sica,
  Universidad de Valencia, 46100 Burjassot (Valencia), Spain}
\address{$^{3}$ Department of Physics, Aristotle University of Thessaloniki,
  Thessaloniki 54124, Greece }
\ead{miga@mpa-garching.mpg.de}
\date{\today}

%\maketitle
\begin{abstract}
We extend a general-relativistic ideal magneto-hydrodynamical code to include the effects of elasticity. Using this numerical tool we analyse the magneto-elastic oscillations of highly magnetised neutron stars (magnetars). In simulations without magnetic field we are able to recover the purely crustal shear oscillations within an accuracy of about a few per cent. For dipole magnetic fields between $5\times10^{13}$ and $10^{15}\,$G the Alfv\'en oscillations become modified substantially by the presence of the crust. Those quasi-periodic oscillations (QPOs) split into three families: Lower QPOs near the equator, Edge QPOs related to the last open field line and Upper QPOs at larger distance from the equator. Edge QPOs are called so because they are related to an edge in the corresponding Alfv\'en continuum. The Upper QPOs are of the same kind, while the Lower QPOs are turning-point QPOs, related to a turning point in the continuous spectrum.
\end{abstract}

\section{Introduction}
The theoretical framework of relativistic elasticity has been developed recently in a series of papers by \cite{Karlovini2003, Karlovini2004, Karlovini2004b, Karlovini2007}. One natural application of this theory is the crust of a neutron star, where the structure of the matter is crystalline and it is able to support shearing motions. When calculating the oscillatory modes supported by the crust, one has to include the effects of gravitation, which influence the theoretically obtained frequencies for isolated neutron stars significantly. Candidates for observations to test theoretical models basing on relativistic elasticity can be found in the decaying tail of a giant burst of a soft-gamma repeater (SGR). In the giant flares of two such objects, SGR 1900+14 and SGR 1806-20, a number of long-lasting, quasi-periodic oscillations (QPOs) have been observed (see \cite{Israel2005} and \cite{Watts2007} for recent reviews). 

The first attempts to explain those QPOs are based on models of purely crustal shear oscillations of an isolated neutron star (see e.g. \cite{Duncan1998, Strohmayer2005, Piro2005, Sotani2007, Samuelsson2007, Steiner2009}). However, SGRs are believed to posses ultra strong magnetic fields $B\approx10^{15}\,$G, and it is necessary to construct self-consistent models including the interaction of the Alfv\'en oscillation with the crustal shear modes (\cite{Levin2006,Glampedakis2006,Levin2007,Lee2007,  Lee2008}). 
A second approach to understand these QPOs is therefore based on purely Alfv\'en oscillations without a crust (see e.g. \cite{Sotani2008,Cerda2009,Colaiuda2009}). In those studies two families of QPOs were found. They are related to the open field lines near the pole and to the closed field lines near the equator. That Alfv\'en QPO model is very attractive, because it
reproduces the near-integer-ratios of the observed 30, 92 and 150\,Hz
frequencies in SGR 1806-20.  The results of the numerical simulation
agree with a semi-analytic model based on standing waves in the
short-wavelength limit \cite{Cerda2009}.

Naturally any realistic model has to include both contributions, the crust and the magnetised core.
In first simplified models, it was shown that the effect of the coupling between crust and core may lead to an absorption of shear modes into a MHD continuum of Alfv\'en oscillations \cite{Levin2007}. Additionally long-lived QPOs may still appear at the turning
points or edges of the continuum.

Here we extend a previous model of \cite{Cerda2009} to simulate coupled, magneto-elastic oscillations in a general-relativistic framework. We use a dipolar magnetic field and a tabulated equation of state (EOS) for dense matter. The numerical simulations are based on state-of-the-art
Riemann solver methods for both the interior MHD fluid and the crust.

We use units where $c=G=1$ with $c$ and $G$ being the speed of light
and the gravitational constant, respectively. Latin (Greek) indices
run from 1 to 3 (0 to 3).

\section{Theoretical model}
Following the analysis by \cite{Cerda2009}, who considered purely Alfv\'en oscillations of magnetars with a two-dimensional, general-relativistic, ideal magnetohydrodynamic code called MCoCoA \cite{Cerda2008}, we extend this code by including the effects of the crust in general relativity.
As in \cite{Carter2006} this can be done by including an additional term $T^{\mu\nu}_\mathrm{elas}$ in the stress-energy tensor
\begin{eqnarray}\label{energytensor}
 T^{\mu\nu} &=& T^{\mu\nu}_{\mathrm{fluid}} + T^{\mu\nu}_{\mathrm{mag}} +
 T^{\mu\nu}_{\mathrm{elas}} \nonumber\\
 &=& \rho h u^\mu u^\nu + P g^{\mu\nu} +  b^2 u^\mu u^\nu + \frac{1}{2} b^2
     g^{\mu\nu} - b^\mu b^\nu - 2 \mu_{\mathrm{S}} \Sigma^{\mu\nu}\label{T^munu} \, ,
\end{eqnarray}
where $\rho$ is the rest-mass density, $h$ the specific enthalpy, $P$
the isotropic fluid pressure, $u^\mu$ the 4-velocity of the fluid,
$b^\mu$ the magnetic field measured by a co-moving observer (with
$b^2:=b^\mu b_\mu$), $\Sigma^{\mu\nu}$ the shear tensor, and
$\mu_{\mathrm{S}}$ the shear modulus, respectively. The latter is
obtained according to \cite{Sotani2007}.
As in the previous work we apply a number of simplifications: (i) a zero temperature
EOS, (ii) axisymmetry, (iii) a purely poloidal magnetic field configuration, (iv) the Cowling approximation, (v) a spherically symmetric background and (vi) small amplitude oscillations. Assumptions (ii) and (iii) lead in the linear approximation to the decoupling of polar oscillations from axial ones. 
We further assume a conformally flat metric 
% %
% \footnote{This provides a very good approximation as our neutron star
%   models are almost perfectly spherically symmetric except for very
%   small deviations due to the presence of an axisymmetric magnetic
%   field.}
% %
\begin{equation}
 ds^2 = - \alpha^2 dt^2 + \phi^4 \left( dr^2 + r^2 d\theta^2 
        + r^2 \sin{\theta}^2 d\varphi^2 \right) \, ,
\end{equation}
where $\alpha$ is the lapse function and $\phi$ the conformal factor.
Employing the induction equation for the magnetic field, the equations of the conservation of energy and momentum can be cast into a conservation law of the following form
\begin{equation}
 \frac{1}{\sqrt{-g}} \left( \frac{\partial\sqrt{\gamma} \bf U }{\partial t} +
 \frac{\partial \sqrt{-g} {\bf F}^i}{\partial x^i} \right) = 0 \, , 
\label{conservationlaw}
\end{equation}
where $g$ and $\gamma$ are the determinants of the 4-metric and 3-metric,
respectively. The two-component state and flux vectors are given by
\begin{eqnarray}
 {\bf U}   &=& [S_\varphi ,\, B^\varphi]  
\label{reduced_withcrust1} \, ,
 \\
 {\bf F}^r &=& \left[ - \frac{b_\varphi B^r}{W} - 2 \mu_S
                      \Sigma^r_{~\varphi} ,\, - v^\varphi B^r
               \right]  
\label{flux_r} \,  ,
\\
 {\bf F}^\theta &=& \left[ - \frac{b_\varphi B^\theta}{W}- 2 \mu_S
                          \Sigma^\theta_{~\varphi} ,\, -v^\varphi
                          B^\theta 
                   \right] \, ,
\label{flux_theta}  
\end{eqnarray}
where $B^i$ are the magnetic field components as measured by an {\it
  Eulerian observer} \cite{Anton2006}, and $W=\alpha u^t$ is the
Lorentz factor.  The shear tensor $\Sigma^{i \varphi}= 1/2 g^{ii}
\xi^\varphi_{~,i}$ contains the spatial derivatives (denoted by a
comma) of the fluid displacement $\xi^\varphi$ due to the
oscillations, which are related to the fluid 4-velocity according to $
\xi^\varphi_{~,t} =\alpha v^\varphi = {u^\varphi} / {u^t}$, where
$v^\varphi$ is the $\varphi$-component of the fluid 3-velocity. 

The boundary conditions at the surface $\xi^\varphi_{~,r}=0$ is a consequence of the continuous traction condition and vanishing currents at the surface of the star. At the crust-core interface the continuous traction condition together with condition of continuous parallel electric field (and so continuous $\xi_\varphi$) imply $\xi^\varphi_{\mathrm{core},r} = \left( 1 +
\delta \right) \xi^\varphi_{\mathrm{crust},r}$ with $\delta = \mu_\mathrm{S}/(b_r b^r)$\,.

The equilibrium models for our simulations are constructed using the LORENE library ({\tt www.lorene.obspm.fr}). 

As an additional tool for the analysis of the coupled magneto-elastic oscillations of the neutron star, 
we extend the semi-analytic model of \cite{Cerda2009} by
including a description of crust-core coupling. That provides a
comparison aiding the interpretation of our numerical results. 
In the linear regime and in the limit of short wave lengths an Alfv\'en wave travels along the magnetic field line corresponding to
\begin{equation}
 \frac{d\mathbf{x}}{dt}=\mathbf{v}_a (\mathbf{x})\,,
\end{equation}
where $\mathbf{v}_a$ is the Alfv\'en velocity.
Assuming standing waves along the magnetic field lines
one can derive the following dispersion relation 
\begin{equation}
 f=\frac{\kappa}{2\pi t_\mathrm{tot}}\,,
\end{equation}
where $t_\mathrm{tot}$ is twice the total travel time of an Alfv\'en wave travelling along a magnetic field line starting from the equatorial plane and ending at the surface or another
point at the equatorial plane. At this point the frequency of the oscillations is completely determined by the magnetic field topology and the boundary conditions. Each field line is oscillating independent from the others with its own frequency. Therefore the collective of all lines  forms a continuum of frequencies.

This model may be modified in the presence of the crust in two limiting cases.
One may assume that the standing wave gets reflected at the crust-core interface, meaning that $t_\mathrm{tot}$ describes twice the total travel time up to the crust-core interface instead of the surface of the star. Simultaneously the boundary condition at the crust-core interface changes to zero amplitude in the crust. This possibility should be valid for low magnetic fields. 

In the second case one may integrate up to the surface of the neutron star, but additionally take the shear velocity in the crust $\mathbf{v}_s$ into account. The resulting travel time is then obtained from
\begin{equation}
 \frac{d\mathbf{x}}{dt}=\sqrt{\mathbf{v}^2_a (\mathbf{x})+\mathbf{v}^2_s (\mathbf{x})}\,.
\end{equation}
This is only an approximation because we are dealing with isotropic shear and so there is no preferred direction in contrast to the case with only magnetic field. Nevertheless it is expected to be valid approximately when the magnetic field dominates in the above equation.

\section{Results}
To check the reliability of our method we first compare the frequencies of the purely shear oscillations obtained in our simulations with those available in the Literature \cite{Sotani2007}. The numerical grid for the simulations consisted of 60 points in $\theta$ direction between 0 and $\pi/2$ and 120 in radial direction from 0 to the surface of the star, corresponding to about 20 points in the crust. The comparison can be found in table\,\ref{tab_shear}. Despite the different methods, i.e. eigenmode calculation in \cite{Sotani2007} and Fourier analysis of the numerical evolution in this work, we are able to recover the frequencies of the crustal shear modes with very good accuracy of at least a few per cent.
\begin{table}
 \caption{\label{tab_shear} Crustal shear frequencies of different equilibrium models described in \cite{Sotani2007}. The values in parenthesis are reference values given in \cite{Sotani2007}.}
\begin{center}
 \begin{tabular}{c|c c c |c}
  \br
Model& \multicolumn{4}{c}{frequency in Hz for mode}\\ \mr
&\multicolumn{3}{c}{n=0 ($\pm$1Hz)}& n=1\\
&l=2&l=3&l=4&$\pm20$Hz\\ \br
APR+DH 1.6 & 23.5 (23.4) & 37.1 (37.0) & 49.8 (49.6) & 880 (860)\\
APR+DH 2.0 & 21.9 (21.3) & 35.1 (33.6) & 46.8 (45.1) & 1070 (1083)\\
L+DH 1.6 & 20.5 (20.6) & 32.5 (32.5) & 43.8 (43.7) & 590 (586)\\
L+DH 2.0 & 19.0 (18.9) & 30.2 (29.9) & 40.5 (40.2) & 720 (713)\\
WFF3+DH 1.6 & 25.2 (25.2) & 39.8 (39.9) & 53.4 (53.5) & 1130 (1101)\\ \br
 \end{tabular}
\end{center}
\end{table}

For the remainder of this section we will present results which refer to the specific stellar model with $1.4\,\mathrm{M_\odot}$  and circumferential radius of about $12.26\,$km. The employed equation of state of the core APR is described by \cite{Akmal1998} and is matched to the one of DH \cite{Douchin2001} for the crust. In any case we do not obtain qualitatively different results for different EOS or the same EOS but different masses of the model.

In the simulations we find three different regimes, depending on the relative
strength between the magnetic field and the shear terms of the crust. For low
magnetic fields $B<5\times10^{13}\,$G the purely crustal shear modes are
recovered as shown above. With increasing field strength, the influence of the
magnetic field increases and the crustal shear modes become damped
\cite{Gabler2010letter}. For larger magnetic fields $B\gtrsim10^{15}\,$G the
magnetic field dominates the evolution and our results approach those of
\cite{Cerda2009}. 

\begin{figure}
 \begin{center}
  \includegraphics[width=\textwidth]{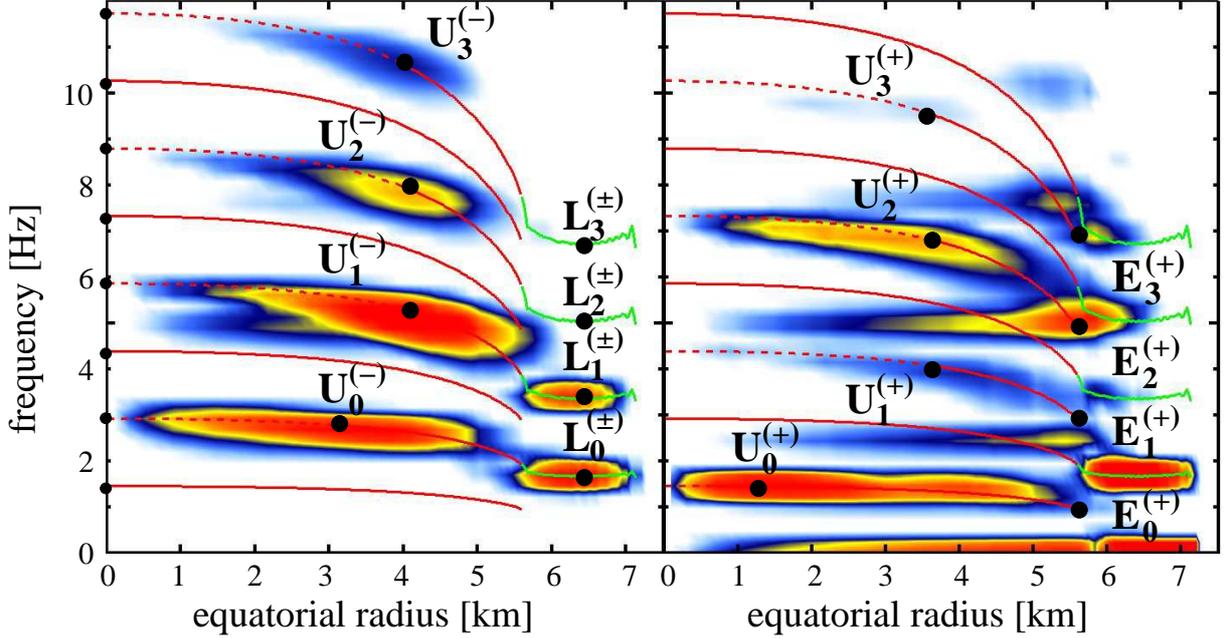}
 \end{center}
\caption{\label{fig_FFT} The magnitude of the Fourier transformation averaged along individual field lines denoted by their crossing point with the equatorial plane. \emph{left} panel: simulation with odd parity; \emph{right} panel: simulation with even parity.  The solid and dashed lines represent the Alfv\'en continuum in the core, obtained with a semi-analytic model. Black points indicate the locations of Upper QPOs $U_n^{(\pm)}$, Lower QPOs $L_n^{(\pm)}$ and Edge QPOs $E_n^{(+)}$. The color scale reaches from white (minimum) to orange-red (maximum).}
\end{figure}

Here we will focus on the intermediate regime, where the crust has a significant
influence on the long-term oscillations occurring in the core of the neutron
star. For magnetic fields between $5\times10^{13}$ and $10^{15}\,$G the coupled
magneto-elastic oscillations hardly reach the surface of the star. Their largest
oscillation amplitudes are located inside the liquid core of the star, see
\cite{Gabler2010letter} for more details. Further changes compared to the case
without crust \cite{Cerda2009}, can be studied in figure \ref{fig_FFT}, where we
show the magnitude of the Fourier transformation averaged along individual
magnetic field lines denoted by their crossing point with the equatorial plane.
The two panels show simulations with either odd (left) or even (right) symmetry
with respect to the equatorial plane. The solid and dashed, red and green lines
indicate the position of the Alfv\'en continuum obtained with the semi-analytic
model, if reflection at the crust-core interface is assumed. 
First we note that the structure with the lowest frequency, the fundamental QPO $U_0^{(+)}$, has even parity. This is in contrast to the simulations without crust, where this particular QPO even did not exist at all and the fundamental oscillation had odd parity. The local maximums in figure \ref{fig_FFT} denoted by $U_n^{(\pm)}$ are directly related to the continuum (red lines). We therefore interpret them as similar to the \emph{Upper} QPOs in \cite{Cerda2009} with the difference that in the presence of a crust they are located significantly away from the polar axis.
Probably the effects of the crust interfere in the building of standing waves near the pole, such that no stable standing waves can be maintained there. We indicate the corresponding region by the dashed lines in figure \ref{fig_FFT}. The second family, called \emph{Lower} QPOs $L_n^{(\pm)}$ around the closed field lines is reproduced qualitatively as in the case without crust. The only difference is that those QPO are in the present case confined to the field lines which close inside the liquid core of the neutron star. This is not surprisingly, because the crust is not expected to influence the oscillations along those magnetic field lines.
In contrast to \cite{Sotani2008, Cerda2009} we find a third family of QPOs related to the last magnetic field line, which just fails to close inside the core. QPOs at similar locations have been found by \cite{Colaiuda2009} in simulations without crust.

The different families can be interpreted according to \cite{Levin2007} in the following way:
the Lower QPOs are related to a turning point in the spectrum (green lines in figure \ref{fig_FFT}; obtained with the semi-analytic model), so they are turning-point QPOs; the Upper QPOs, which were related in \cite{Cerda2009} to the turning point at the polar axis change their character and become edge QPOs located at the edge of the continuum indicated by the end of the solid and the beginning of the dashed lines; the new family of even QPOs may be related to those parts of the continuum of the open field lines, which do not connect to that of the closed field lines. Since they are located at an edge of the continuum (solid red lines), we call this new family \emph{Edge} QPO $E_n^{(+)}$.

\section{Conclusions}

At magnetic field strength below $5\times10^{13}\,$G the torsional oscillations in magnetars are dominated by the crustal shear modes. For very large fields $B>10^{15}\,$G our results approach those of \cite{Cerda2009} without crust. In the intermediate regime the crustal modes disappear quickly, but the presence of the crust influences the magneto-elastic oscillations significantly in different ways:
First, the structure of the obtained Upper QPOs changes as a consequence of a reflection at the crust-core boundary. Those QPOs are now located between the polar axis and the equator. Additionally they have a minimum at the crust-core interface and vanishing amplitude inside the crust; second a new family of Edge QPOs appears due to the coupling of the field lines which just fail to close inside the liquid core; third the Lower QPOs are limited to the field lines closing inside the liquid core.

To summarize the shear oscillations of the crust are damped efficiently and the magneto-elastic QPOs have negligible amplitudes in the crust for magnetic field strengths in the range of $5\times10^{13}$ to $10^{15}\,$G. Therefore, our simulations favor magneto-elastic oscillations at field strength $>10^{15}\,$G as a possible explanation for the observed magnetar QPOs.

% 
% \begin{figure}
%  \begin{center}
%   \includegraphics[width=\textwidth]{overlap.eps}
%  \end{center}
% \caption{\label{fig_damping} damping of crustal mode}
% \end{figure}

\section*{References}
\bibliographystyle{iopart-num}
\bibliography{magnetar}

\end{document}